\documentclass[twocolumn]{aastex63}

\shorttitle{MRG-M0138}
\shortauthors{Jafariyazani et al.}

\graphicspath{{./}{figures/}}

\begin{document}

\title{Resolved Multi-element Stellar Chemical Abundances in the Brightest Quiescent Galaxy at \textit{z} $\sim$ 2}

\author[0000-0001-8019-6661]{Marziye Jafariyazani}
\affil{The Observatories of the Carnegie Institution for Science, Pasadena, CA, USA }
\affil{Department of Physics and Astronomy, University of California, Riverside, CA, USA }

\author[0000-0001-7769-8660]{Andrew B. Newman}
\affil{The Observatories of the Carnegie Institution for Science, Pasadena, CA, USA }

\author{Bahram Mobasher}
\affil{Department of Physics and Astronomy, University of California, Riverside, CA, USA }

\author{Sirio Belli}
\affil{Center for Astrophysics | Harvard and Smithsonian, Cambridge, MA, USA}

\author{Richard S. Ellis}
\affil{University College London, London, UK}

\author{Shannon G. Patel}
\affil{The Observatories of the Carnegie Institution for Science, Pasadena, CA, USA }

\begin{abstract}

Measuring the chemical composition of galaxies is crucial to our understanding of galaxy formation and evolution models. However, such measurements are extremely challenging for quiescent galaxies at high redshifts, which have faint stellar continua and compact sizes, making it difficult to detect absorption lines and nearly impossible to spatially resolve them. Gravitational lensing offers the opportunity to study these galaxies with detailed spectroscopy that can be spatially resolved. In this work, we analyze deep spectra of MRG-M0138, a lensed quiescent galaxy at \textit{z} = 1.98 which is the brightest of its kind, with an H-band magnitude of 17.1. Taking advantage of full spectral fitting, we measure $[{\rm Mg/Fe}]=0.51\pm0.05$, $[\rm{Fe/H}]=0.26\pm0.04$, and, for the first time, the stellar abundances of 6 other elements in this galaxy. We further constrained, also for the first time in a $z\sim2$ galaxy, radial gradients in stellar age, [Fe/H], and [Mg/Fe]. We detect no gradient in age or [Mg/Fe] and a slightly negative gradient in [Fe/H], which has a slope comparable to that seen in local early-type galaxies.
Our measurements show that not only is MRG-M0138 very Mg-enhanced compared to the centers of local massive early-type galaxies, it is also very iron rich. These dissimilar abundances suggest that even the inner regions of massive galaxies have experienced significant mixing of stars in mergers, in contrast to a purely inside-out growth model. The abundance pattern observed in MRG-M0138 challenges simple galactic chemical evolution models that vary only the star formation timescale and shows the need for more elaborate models.

\end{abstract}


\section{Introduction} \label{sec:intro}

\begin{figure*}
\includegraphics[width=1\textwidth]{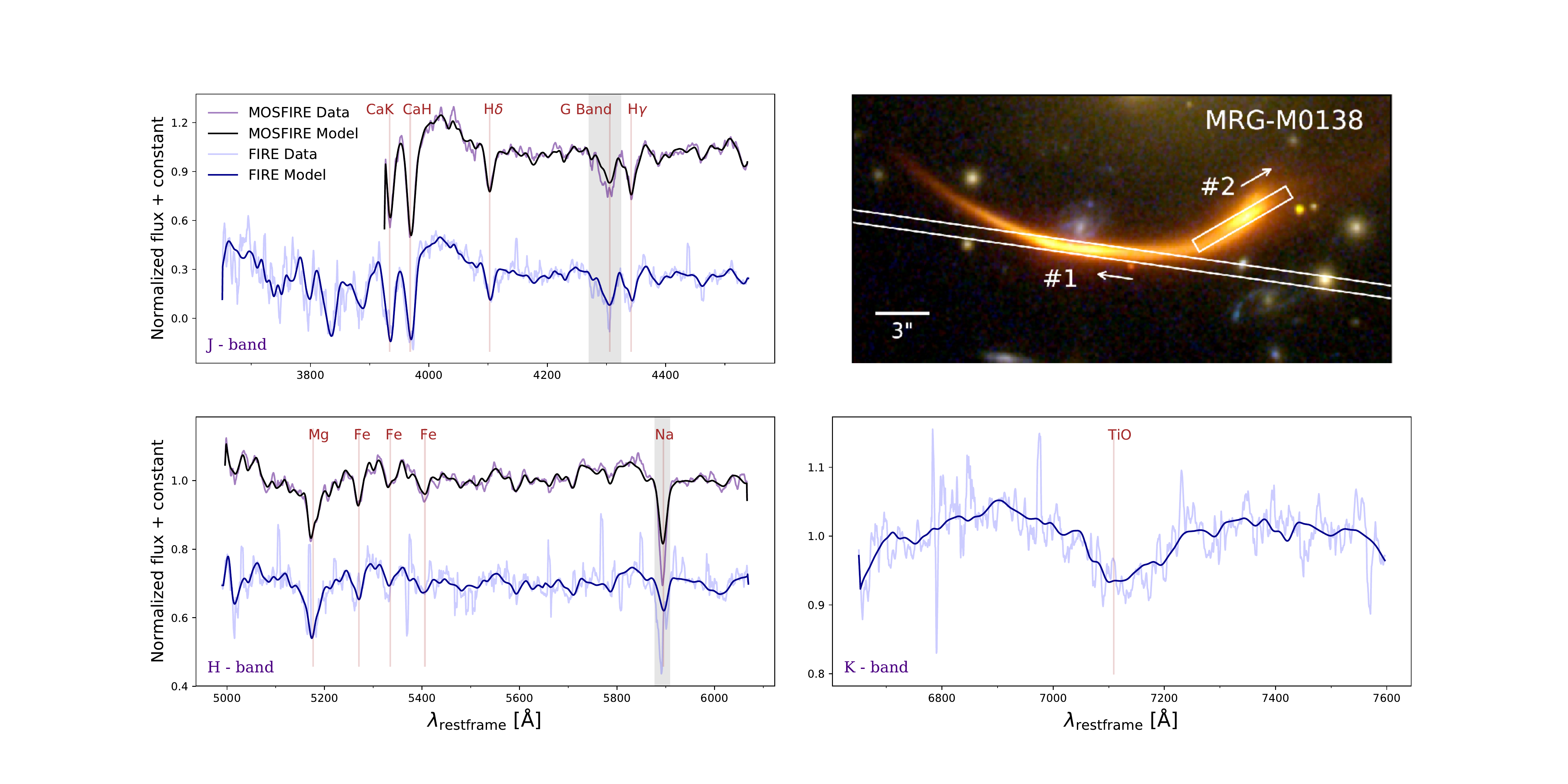}
\caption{The upper right panel presents \emph{HST} observations of the two brightest images of MRG-M0138 along with the placement of the slit on each image (reproduced from \citealt{Newman2018/2}). Other panels show the integrated spectra of these two images observed with Keck/MOSFIRE (purple, Image 1 in the upper right panel) and Magellan/FIRE (light blue, Image 2) in J, H, and K bands. The best-fit model to each spectrum is shown in solid black or blue. Grey shaded regions are masked in the spectral fitting process. For display purposes, all spectra are smoothed to 370 km s$^{-1}$ by taking the inverse variance weighted mean.}
\label{fig:data+model}
\end{figure*}

Chemical abundances in early-type galaxies have been one of the main tools for studying their formation history. Specifically the $\alpha$-enhancement ($\alpha$-element-to-iron abundance ratio compared to the solar abundance ratio, e.g., [Mg/Fe]), is sensitive to the timescale of star formation because of the different lifetimes of the stars that produce these elements in supernovae (SNe). SN Type II (core-collapse) release mostly $\alpha$-elements including oxygen, magnesium, silicon, etc. (\citealt{Woosley1995}) whereas SN Type Ia mostly release iron and other iron peak elements. This implies that galaxies with more extended star formation histories (SFHs) are expected to have a lower [$\alpha$/Fe] and higher [Fe/H].

The $\alpha$-enhancement, as a timescale diagnostic, has been extensively used to infer the formation histories of early-type galaxies in the local universe (e.g., \citealt{Thomas2003,Greene2019}). However, these archaeological studies can reconstruct the formation history of the stars currently in the galaxy, and if there has been a significant amount of merging, then those stars were born in a range of different progenitor galaxies with potentially different formation histories. Therefore, to study conditions in the main progenitors of today’s ellipticals, including highly star-forming dust-obscured galaxies at $\textit{z} > 3$ (e.g., \citealt{Toft2014_SMG,Casey,valentino2020}), we need to observe quiescent galaxies closer to the epoch of quenching, before the majority of merger activity had occurred and polluted the population. At high redshifts, nearly all chemical studies have focused on the gas-phase metallicity, measured using nebular emission lines, which traces the current composition of the interstellar medium (ISM) and is affected by inflows and outflows. Stellar abundances requiring absorption line measurements are more challenging, but they provide complementary information by tracing the composition of ISM when the stars were forming. Furthermore, stellar abundances are the only chemical probe available for quiescent galaxies that lack bright emission lines and may be gas-poor. However, the faintness of high-\textit{z} quiescent galaxies makes spectroscopic observations extremely challenging.

In recent years, several authors have attempted to extend stellar chemical abundance studies to higher redshifts using a single or stacked spectra of a handful of quiescent galaxies (e.g., \citealt{Onodera2015,Lonoce2015,Kriek2019,Lonoce2020}). Currently, there is only one quiescent galaxy at \textit{z} $\sim$ 2 for which [Fe/H] and [Mg/Fe] have been measured to $\sim$ 0.1 dex precision (COSMOS-11494, \citealt{Kriek2016}). This galaxy is the most Mg-enhanced massive galaxy ever observed and is thus believed to have had a very short star formation timescale compared to that inferred from local samples. Results from this unique object suggested that the chemical composition of massive galaxies may have evolved substantially after they quenched, likely through mergers. However, more observations are needed to confirm this scenario.

In this work, we analyze the spectra of MRG-M0138, a gravitationally lensed quiescent galaxy at \textit{z} = 1.98 discovered by \cite{Newman2018a}. We measure its detailed stellar abundance pattern and compare it to COSMOS-11494 and local massive galaxies that are representative of its descendant. Moreover, for the first time for such a distant quiescent galaxy, we measure the gradient in age, $\alpha$-enhancement and iron abundance to investigate the uniformity of the star formation history. In Section \ref{sec:data} we briefly present our data, in Section \ref{sec:spectral fitting} we explain the full spectral fitting process, models and assumptions, and in Section \ref{sec:results} we present our results, the measured age and chemical abundances of 8 elements. We discuss the implication of our results in Section \ref{sec:discussion}. Where necessary we assume a flat $\Lambda$CDM cosmology with $\Omega_m = 0.3$ and $H_0 = 70$~km~s${}^{-1}$.

\section{Data} \label{sec:data}

\begin{figure*}
\includegraphics[width=1\textwidth]{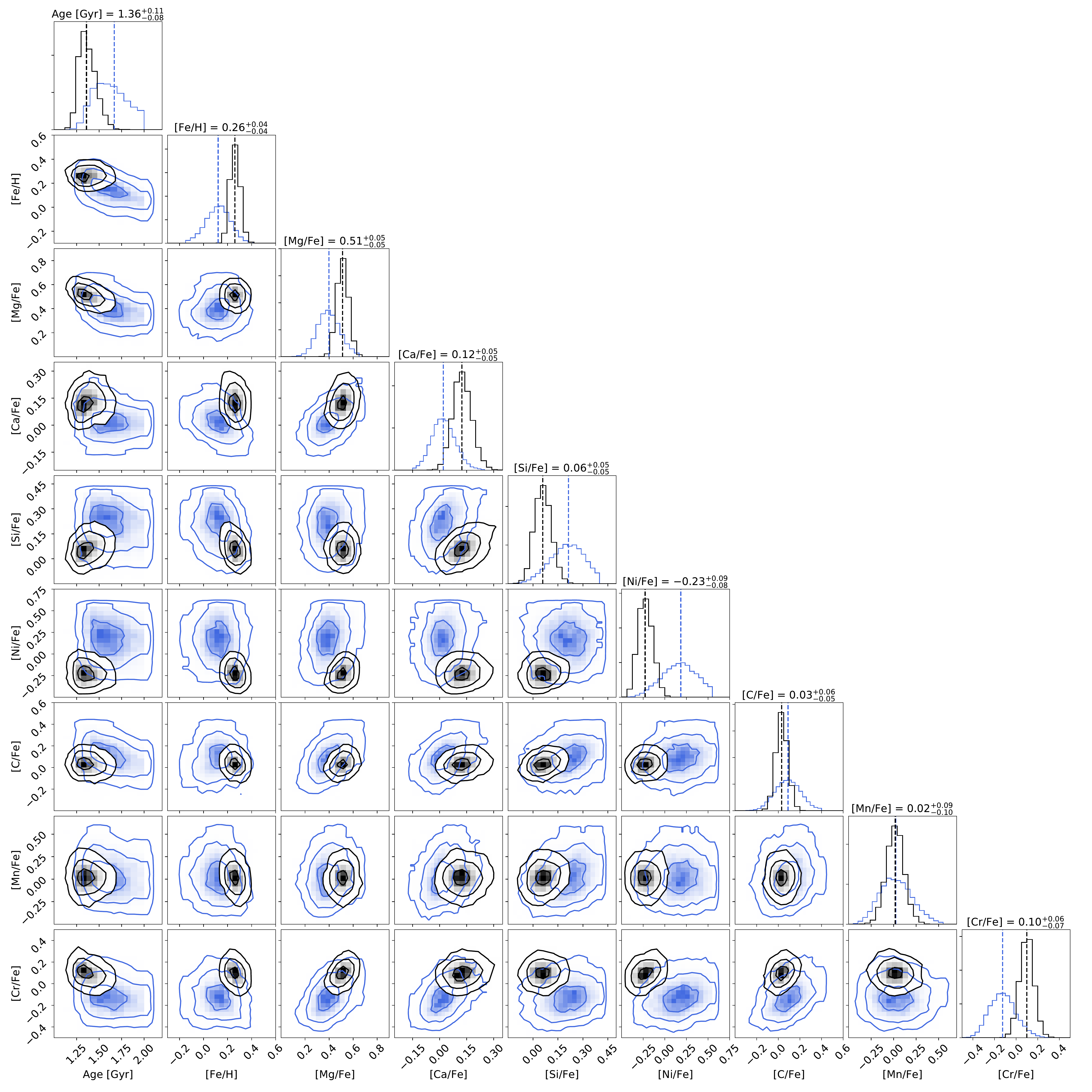}
\caption{Corner plot showing covariances among age, [Fe/H], and the abundances of elements for which [X/Fe] is measured to $ <0.1$ dex precision in the MOSFIRE spectrum. Black contours show results from fitting the MOSFIRE spectrum, and blue contours correspond to fitting the FIRE spectrum, which has lower S/N. These contours correspond to 1$\sigma$, 2$\sigma$ and 3$\sigma$ levels. Titles show the median of the marginalized posteriors obtained from the MOSFIRE spectrum }
\label{fig:cornerplot}
\end{figure*}

 MRG-M0138 is a gravitationally lensed galaxy with five different images. Two spectroscopic observations have been made of this galaxy, one of its brightest image (\textit{H}=17.1) taken with the near-infrared spectrograph MOSFIRE (\citealt{McLean2010_MOSFIRE,MOSFIRE}) at the Keck 1 telescope in the J and H bands, and the other of the second-brightest image taken with the near-infrared echellete FIRE (\citealt{FIRE}) at the Magellan Baade telescope. The MOSFIRE spectrum has a signal-to-noise ratio (S/N) of 137 per velocity dispersion element ($\sim$300 km/s) in H band, which is by far the most detailed spectrum of a quiescent galaxy at a similar redshift. The FIRE spectrum covers a wider wavelength range with a lower S/N=77 and is mainly used as a cross-check in this paper. Details about these observations and data reduction processes can be found in \cite{Newman2018a}. 
 
 We also have resolved spectra in a series of seven bins along the MOSFIRE slit, which we use to derive age and metallicity gradients across this object (for details about these spectra see \citealt{Newman2018/2}). Also, using the resolved rotation curve of this galaxy (\citealt{Newman2018/2}), we are able to remove rotation from the two-dimensional spectrum. After this derotation, the galaxy spectral features are aligned vertically while the sky emission lines are tilted. By extracting a one-dimensional spectrum using inverse variance weighting, we then obtain a remarkably clean integrated spectrum that is virtually free of the sky line residuals that normally affect near-infrared spectra.
 
 The wavelength ranges used for the fitting process in this work are 3925 - 4540 \AA \space and 4995 - 6070 \AA \space for the MOSFIRE spectrum and 3650 - 4540 \AA, 4990 - 6070 \AA \space and 6650 - 7600 \AA \space for the FIRE spectrum. In both spectra, we mask the region between 4270 and 4325 \AA \space where there is evidence for residual telluric absorption. We also mask the Na D line, which we found to be too strong to be a purely stellar feature and thus requires additional absorption by an interstellar medium (see Section \ref{subsec: NaD}). The two brightest images of the MRG-M0138 and our integrated spectra are shown in Figure \ref{fig:data+model}.
 
 \section{Full spectral fitting} \label{sec:spectral fitting}
 We analyze our spectra using \textit{alf} (\citealt{Conroy2012a,conroy2018}), which is an absorption line fitter of near-infrared and optical spectra of stellar systems older than 1 Gyr. This code uses a Markov Chain Monte Carlo (MCMC) algorithm to fit the data to combined libraries of empirical and synthetic stellar spectra covering a wide range of parameter space (\citealt{Sanchez-Blazquez2006,choi2016,Villaume2017}). We run \textit{alf} in full mode, which fits for 46 parameters as described by \cite{conroy2018}, including velocity, velocity dispersion, age, abundances of O, C, N, Na, Mg, Si, K, Ca, Ti, V, Cr, Mn, Co, Ni, Cu, Sr, Ba and Eu, along with various nuisance parameters. We note that the S/N of our spectra is not sufficient to meaningfully constrain all of the abundance ratios, but we identify 8 elements whose abundance is measured to $<$ 0.1 dex precision from the MOSFIRE spectrum and confine our analysis in the rest of the paper to these species: Fe, Mg, Ca, Si, Ni, C, Mn and Cr. We should note that although the MOSFIRE spectrum formally constrained the Ti abundance to 0.1 dex precision, the spectrum does not cover the strong TiO absorption features, so we exclude it from our results. In the fitting process, errors are rescaled by a jitter term to ensure they are realistic, which is useful in the near-infrared where systematic errors in the data reduction can be significant.
 
 In our modeling, we assume a single age population and a Kroupa IMF (\citealt{Kroupa}). We allow the abundance of individual elements ([X/H]) to vary from -0.3 to 1. We also perform following robustness tests: changing the degree of the polynomial fitted to the continuum, fixing the abundances of poorly constrained elements to values typical of local massive early-type galaxies (ETGs), allowing the response function, which determine the change in the spectrum due to a change in the abundance of a single element, to vary with metallicity (rather than using response functions calculated at solar metallicity, our default procedure), and restricting the fit to $\lambda > 4000$~\AA~to exclude the Ca HK lines. Results from all of these tests are well within the error-bar of the final reported values.
 
We followed the same procedure for each of the 7 resolved spectra. However, due to their lower S/N, we focus only on the age, [Fe/H], and [Mg/Fe] derived from these spectra.
 
Moreover, in order to compare the properties of our high-\textit{z} quiescent galaxy with local counterparts, we also perform the fitting process with exactly the same models and assumptions on six stacked spectra of SDSS (Sloan Digital Sky Survey) early-type galaxies from \cite{Conroy2014}. These high S/N spectra are from the inner 0.5 R$_{e}$ of passive galaxies, and are binned by their velocity dispersion spanning from 100 to 320 km s$^{-1}$. The only difference is the wavelength range of the fit, which is from 4000 - 7300 \AA \space and 8000 - 8850 \AA \space for SDSS spectra.

\begin{figure*}
\includegraphics[width=1\textwidth]{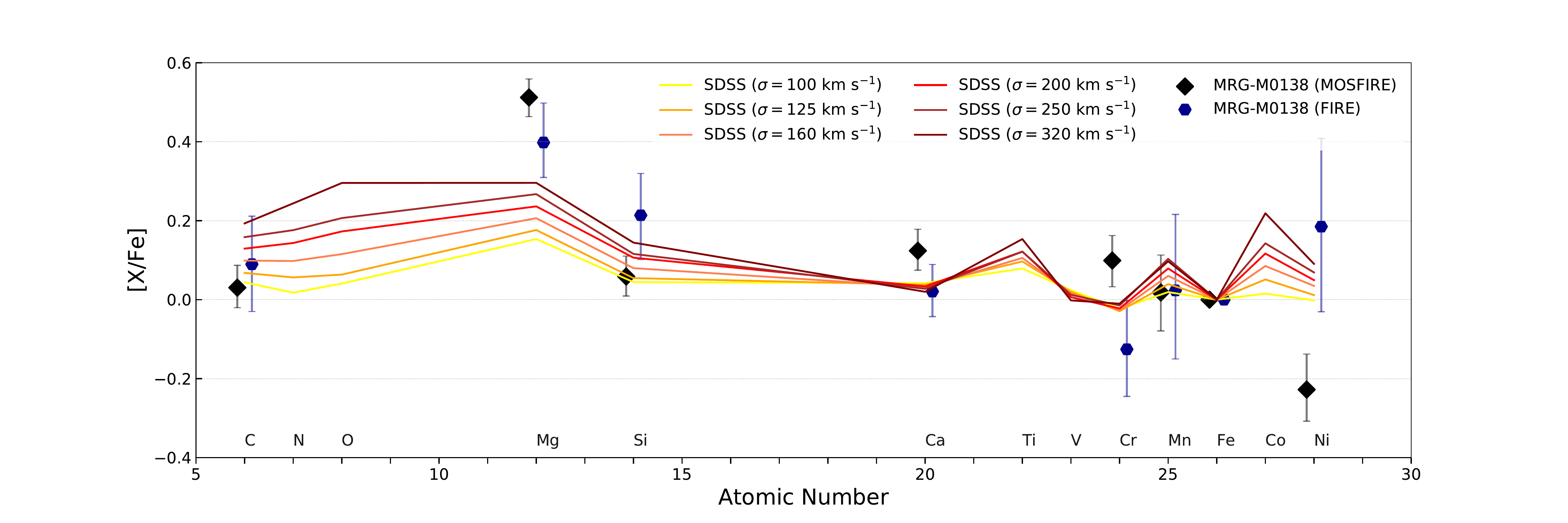}
\caption{[X/Fe] for different elements derived for MRG-M0138 and stacked spectra of SDSS early-type galaxies. Black diamonds represents values for MRG-M0138 derived from the MOSFIRE spectrum, and blue hexagons correspond to the FIRE spectrum. Each colored solid line represents local ETGs with a specific velocity dispersion, $\sigma$, as indicated in the legend. Abundances of most of the elements increase with increasing $\sigma$. Interestingly, [Mg/Fe] is significantly higher in MRG-M0138 than is typical of local galaxies, even those with the highest $\sigma$.}
\label{fig:atomicNumber}
\end{figure*}

\begin{figure*}
\includegraphics[width=1\textwidth]{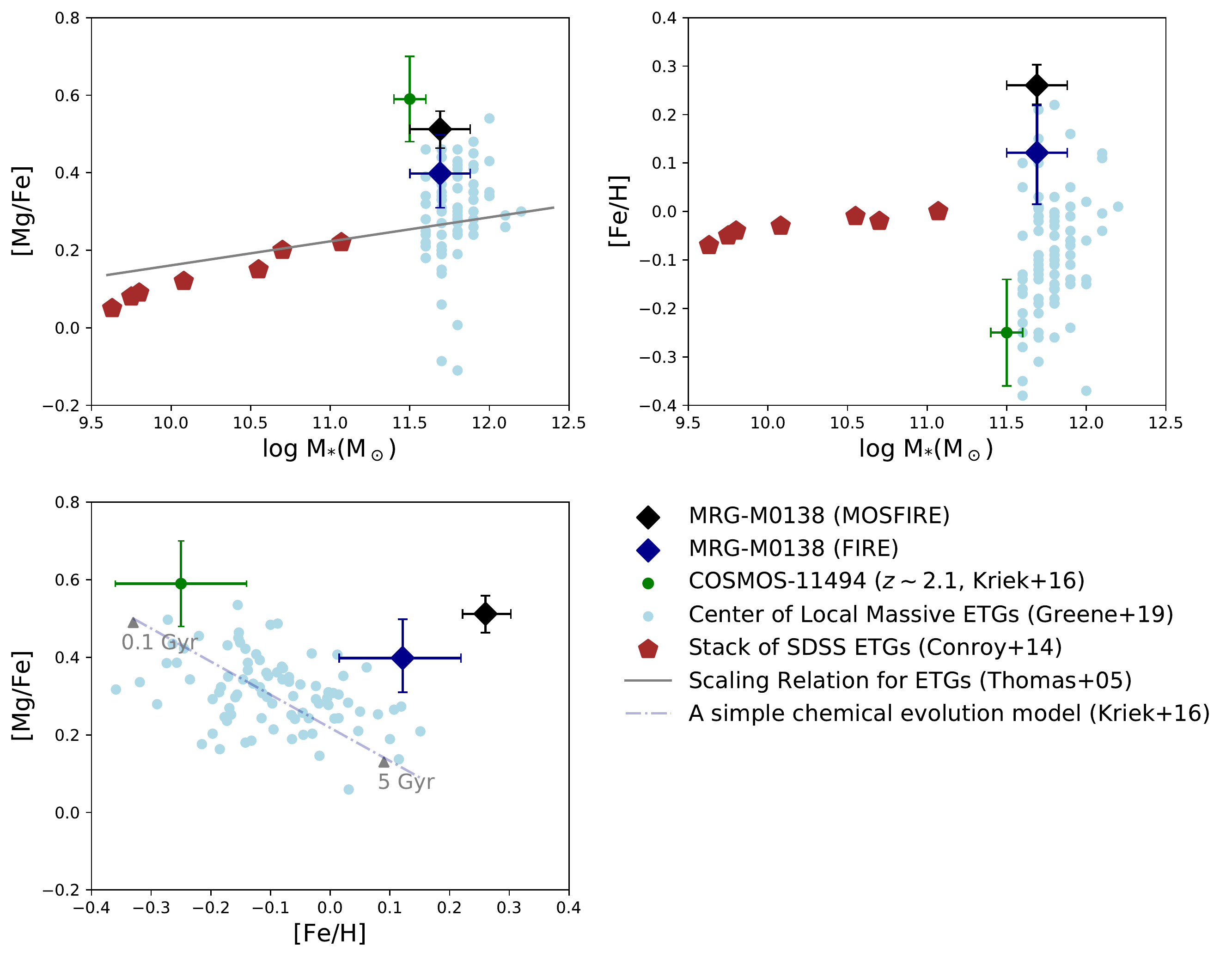}
\caption{Relationships among stellar mass, [Mg/Fe], and [Fe/H] for MRG-M0138 (black and blue diamonds, corresponding to the MOSFIRE and FIRE measurements), COSMOS-11494 at $z=2.1$ (green circle; \citealt{Kriek2016}), and the centers of local massive ETGs (light blue circles; \citealt{Greene2019}). Red pentagons show measurements derived from the SDSS stacks of \cite{Conroy2014}. The solid line in the upper left panel shows the scaling relation from \cite{Thomas2005}. The $\textit{z} \sim$ 2 galaxies are distinct from each other and from the centers of local massive ETGs in the [Mg/Fe] vs. [Fe/H] plane (lower panel). A simple chemical evolution model (dot-dashed line) from \cite{Kriek2016} could plausibly explain the abundances of COSMOS-11494 with a very short formation timescale, but this model cannot account for the high [Mg/Fe] and [Fe/H] of MRG-M0138.}
\label{fig:MgFe}
\end{figure*}

\begin{figure}
\includegraphics[width=0.5\textwidth]{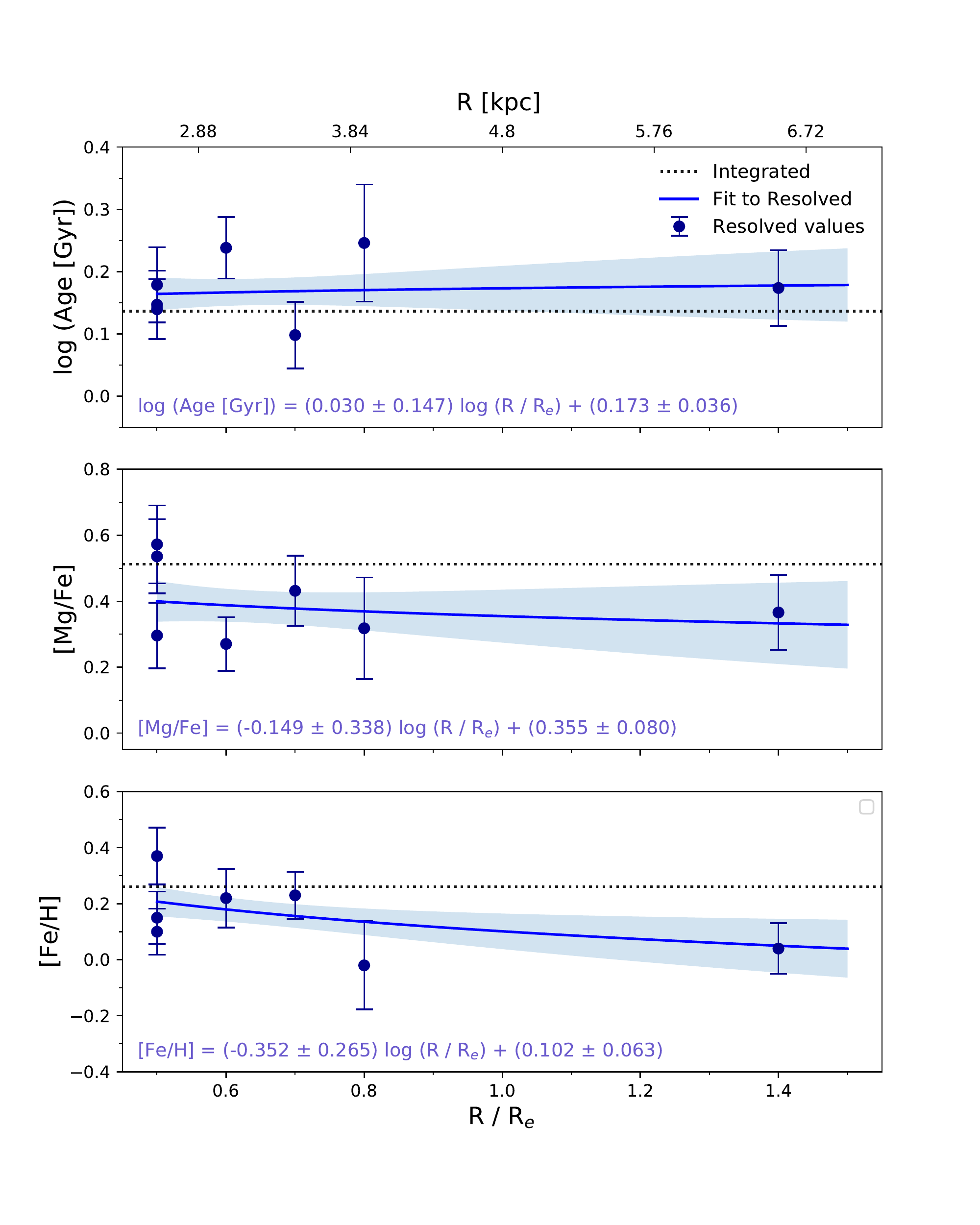}
\caption{Spatially resolved age, [Mg/Fe] and [Fe/H] for MRG-M0138. The flux-weighted radius in each spatial bin is indicated as $R/R_{e}$. Black dotted lines represent the measured value from the integrated spectra. Solid blue line represents the fit to the resolved data in the form of A $\times$ log ($R/R_{e}$) + B, where A and B are presented in each panel. We find that age and [Mg/Fe] are uniform over $0.5-1.4 R_e$, while we detect a marginally negative gradient in [Fe/H].}
\label{fig:resolved}
\end{figure}

\section{Results} \label{sec:results}

Figure \ref{fig:data+model} presents the best-fit models to each spectrum, which provide excellent fits. In Figure \ref{fig:cornerplot}, corner plots show the derived values of age, [Fe/H], and the abundances of 7 elements (Mg, Ca, Si, Ni, C, Mn, Cr) that have $< 0.1$ precision from the MOSFIRE fit, which is unique for a \textit{z} $\sim$ 2 galaxy. Purple contours show results from the MOSFIRE data, and blue contours show results from the FIRE spectrum. Posteriors from the two spectra overlap and we assess the consistency between the two sets of parameters using the Q$_{DM}$ statistic which follows a $\chi^2$ distribution (Equation 49, \citealt{Raveri}). We derive a $p$-value of 0.04 that indicates marginal tension at the $\sim2\sigma$ level between two sets of parameters. This might reflect small residual systematics in the spectra, or the fact that they probe different light-weighted radii in the galaxy due to the lens mapping. We also found that excluding [Ni/Fe] from the comparison removed the tension ($p = 0.13$) and implied full consistency.

The estimated age of the galaxy is 1.37 $\pm$ 0.11 Gyr from the MOSFIRE spectrum (corresponding to $z_{\rm form} = 3.3 \pm 0.2$) and 1.67 $\pm$ 0.23 Gyr from the FIRE spectrum ($z_{\rm form} = 3.8 \pm 0.5$). 

\subsection{Comparison to COSMOS-11494 and local massive ETGs}

We compare the abundances of different elements in this unique high-\textit{z} quiescent galaxy with the chemical abundances of local ETGs in Figure \ref{fig:atomicNumber}. In this figure, solid lines represent [X/Fe] derived from stacked spectra of the inner regions of SDSS early-type galaxies. It is known that the star formation history of passive galaxies is well correlated with their velocity dispersion (e.g., \citealt{Grave2010}), and that is why these galaxies are binned by this parameter. Figure \ref{fig:atomicNumber} shows that the abundance of some elements, such as calcium, does not depend on the velocity dispersion, whereas for some other elements, such as magnesium and carbon, the abundance is a strong function of velocity dispersion (\citealt{Conroy2014}). The abundance ratios of MRG-M0138 and this local sample of ETGs are relatively consistent, except for [Mg/Fe], which is significantly ($\sim$ 0.2 dex) higher for MRG-M0138 even compared to local ETGs at the highest velocity dispersion bin (320 kms$^{-1}$). The only other \textit{z} $\sim$ 2 quiescent galaxy with a comparably precise measurement is COSMOS-11494 (\citealt{Kriek2016}), which also shows a strong Mg-enhancement.

We then present the relation between [Mg/Fe], [Fe/H] and stellar mass (M$_{*}$) for MRG-M0138 and COSMOS-11494 in Figure \ref{fig:MgFe}. To allow for comparison between the stellar populations of quiescent galaxies at the present time and \textit{z} $\sim$ 2, we also show the abundances of a sample of local ETGs from the MASSIVE survey (\citealt{MassiveI}). Considering the high stellar mass of MRG-M0138 (log M$_{*}=11.69\pm$0.12), this galaxy is likely the progenitor of a very massive local elliptical, and the MASSIVE survey is an integral field spectroscopic survey of the most massive early-type galaxies within $\sim$ 108 Mpc, thus providing a perfect comparison sample. Chemical abundances and masses of this sample are adopted from \cite{Greene2019}; we use abundances measured in the inner 2 kpc. Additionally, we added the sample of the stacked spectra of SDSS galaxies from \cite{Conroy2014} to the first two panels of Figure \ref{fig:MgFe}, to show the stellar mass dependence of [Mg/Fe] and [Fe/H] over a wide range of stellar mass. As predicted from local samples, [Mg/Fe] is increasing with stellar mass, but both massive \textit{z} $\sim$ 2 galaxies lie noticeably away from the simple scaling relation based on local ETGs adopted from \cite{Thomas2005}. Even considering the scatter among local ETGs in the MASSIVE sample, COSMOS-11494 has higher ratio of [Mg/Fe] than any of the local ETGs, and MRG-M0138 is comparable to the most Mg-enhanced galaxies in this sample.

In contrast to the relatively similar behavior of MRG-M0138 and COSMOS-11494 in the [Mg/Fe]-M$_{*}$ parameter space, these two galaxies are completely distinct from each other when looking at the [Fe/H]-M$_{*}$ relation. Generally, [Fe/H] is not a strong function of stellar mass, and in the high mass regime, local ETGs from MASSIVE sample span a relatively wide range of [Fe/H], with values as low as $-$0.36 to a maximum of $+$0.15. In this parameter space, MRG-M0138 with [Fe/H] of $+$0.26 $\pm$ 0.04 is more iron-rich than any of these local galaxies, wheres COSMOS-11494 with [Fe/H]= $-$0.25 $\pm$ 0.11 is among the galaxies with the lowest [Fe/H] ratio. This is also emphasized in the last panel of Figure \ref{fig:MgFe}, which shows that both \textit{z} $\sim$ 2 galaxies are distinct from the local sample but with different abundance patterns, which implies significant differences in the enrichment histories of these galaxies. 

Furthermore, the [Ca/Fe] ratio is also significantly different in MRG-M0138 and COSMOS-11494. MRG-M0138 has [Ca/Fe]=$+0.12 \pm 0.05$, comparable to local ETGs, whereas COSMOS-11494 is very Ca-enhanced ([Ca/Fe]=$+0.59 \pm 0.07$; \citealt{Kriek2016}). Its very high [Ca/Fe] ratio is surprising, since local ETGs show almost no Ca-enhancement. In MRG-M0138, on the other hand, the abundance of Ca does not follow Mg, which is consistent with previous studies that showed a different behavior of these two elements, specifically the underabundance of Ca relative to other $\alpha$-elements (e.g., \citealt{Worthey_CaVsMg,Thomas_Ca,Smith_Ca_Mg}). Figure \ref{fig:atomicNumber} shows one aspect of the dissimilar behavior of Mg and Ca: [Mg/Fe] increases with increasing velocity dispersion, whereas [Ca/Fe] is almost constant.

\subsection{Stellar Age and Abundance Gradients}

We present our unique measurement of the spatially resolved age, [Mg/Fe], and [Fe/H] for MRG-M0138 in Figure \ref{fig:resolved}. In each spatial bin, we take the light-weighted radial position from \citet{Newman2018/2}. We fit our resolved data with a functional form of A$\times$log ($R/R_{e}$) + B, where A and B are presented on each panel of Figure \ref{fig:resolved}. We do not detect an age gradient ($A = 0.03 \pm 0.15$ Gyr~dex${}^{-1}$) or a gradient in [Mg/Fe], which suggests that the star formation history was fairly uniform across this galaxy. However, we find a marginally negative gradient for [Fe/H] with a slope of $-0.35 \pm 0.27$. These gradients are comparable to those seen in local early-type massive galaxies. For example, \cite{Greene2015} found gentle gradients in age ($A = -$0.35 $\pm$ 0.11 Gyr dex${}^{-1}$, see their Table 1) and [Fe/H] ($A = -$0.46 $\pm$ 0.11) and no gradient in [Mg/Fe] for the galaxies in the highest velocity dispersion bin ($\sigma > 290$ km~s${}^{-1}$) of the MASSIVE survey.

The small metallicity gradient in MRG-M0138 ($-0.040 \pm 0.028$ dex/kpc) is comparable to the gas metallicity gradients in high-\textit{z} massive star-forming galaxies, and based on simulations, seems to require strong feedback mechanisms during the star-forming phase to mix the gas throughout the disk of this galaxy (e.g., \citealt{Gibson2013,Leethochawalit2016, Ma2017}). Fairly uniform stellar age and a uniformly high [Mg/Fe] imply that star formation must have stopped abruptly and relatively uniformly across the disk around $z\sim3.3$. This is consistent, at least qualitatively, with the short gas depletion times and disky kinematics observed in many high-z submillimeter galaxies proposed as progenitors of galaxies like MRG-M0138 (\citealt{Hodge_smg,Aravena_smg,Jim_smg}).

Another issue of key theoretical importance is the link between quenching and bulge formation. Suggested by compaction models, violent disk instabilities funnel gas into galaxy centers, leading to nuclear star formation just prior to galaxy-wide quenching via gas consumption and outflows (e.g., \citealt{Dekel2014,Zolotov2015,Tacchella2016,Wu2020}). However, some observation of quiescent galaxies at $z\sim2$ reported no indication of a bulge (\citealt{Toft2017_nature}), calling into question the generality of this model. In this context, although MRG-M0138 has a small bulge-like structure resolved in \emph{HST} imaging ($R_e = 800$~pc) that contains 26\% of the light (\citealt{Newman2018a}), our ground-based spectroscopy cannot reach within $0.5 R_e = 2.4$~kpc, which is necessary to investigate the bulge stellar populations and thereby test compaction or other related quenching models. These further investigations will become possible with future \emph{JWST} observations.

\subsection{Evidence for an Interstellar Medium} \label{subsec: NaD}

Na D absorption is very strong in MRG-M0138. If we interpret the absorption as entirely photospheric, our alf fits imply an unreasonably high [Na/H] $\sim$ 2.5. This strongly suggests that an interstellar medium (ISM) is responsible for some of the Na D absorption. To place a lower limit on the column density of ISM, we assume a maximum stellar [Na/H] = $+$1, consistent with observations of local ETGs (\citealt{Conroy2014}), and measure the equivalent width (EW) of Na D for a model with the age of MRG-M0138. In this model, ${\rm EW(Na D)} = 3.57$~\AA, and the measured value in the MOSFIRE spectrum, is ${\rm EW(Na D)} = 5.02 \pm 0.12$~\AA. Now we can simply assume that the excess absorption is due to the ISM, ${\rm EW(Na D)}_{\mathrm{ISM}} = 1.45$~\AA, and then assuming an unsaturated interstellar absorption line, we can estimate a lower limit for the column density of the sodium, ${\rm N(Na)} > 5 \times 10^{12}$~cm${}^{-2}$, using the relation from \cite{Spitzer1978} and the oscillator strength from \cite{Morton_Atomiclines}.

Using the linear relation between the neutral sodium column density and neutral atomic and molecular hydrogen column density N(H) in the diffuse ISM from \cite{Farlet_Na_H}, we can estimate N(H) $\sim$ 9$\times 10^{20}$ cm$^{-2}$. If we assume the neutral hydrogen is distributed within the effective radius ($R_{e,{\rm maj}} = 4.8$~kpc, axis ratio $b/a = 0.26$; \citealt{Newman2018a}) with this column density, we find a lower limit for the neutral hydrogen mass within 1 R$_{e}$ of $\sim$ 10$^{8}$ M$_{\odot}$. This is small compared to the stellar mass of the galaxy ($M_{\rm H I + H_2} / M_* > 3 \times 10^{-4}$), indicating that although quenching did not completely remove the ISM, our observations would be consistent with a very small gaseous reservoir remaining. We note this is a very conservative lower limit on the gas mass of the MRG-M0138, which is far below the current detectable range of the deepest radio observations targeting dust continuum or CO line emission. Therefore, it is complementary to radio observations that detect or place an upper limit on the ISM mass of this galaxy, which would better provide insight into the processes that quenched its star formation.

\section{Discussion} \label{sec:discussion}

We have obtained an extremely high S/N spectrum of MRG-M0138, a lensed quiescent galaxy at \textit{z} = 2. We measured the stellar age, [Fe/H], and abundance ratios [X/Fe] for 7 species for the first time in a \textit{z} $\sim$ 2 galaxy. Interestingly the abundance ratio pattern (Figure \ref{fig:atomicNumber}) seems compatible with massive local ETGs for most elements, with the major exception of Mg. This is striking, since Mg is produced exclusively in core-collapse SNe and so is a pure tracer of massive stars. The very high [Mg/Fe] ratio we find is particularly remarkable when coupled with the high [Fe/H] (Figure \ref{fig:MgFe}), which renders MRG-M0138 dissimilar from the centers of local ETGs.

 The chemical abundances of MRG-M0138 and COSMOS-11494, the only two quiescent galaxies at $z\sim2$ with precise abundance measurements, suggest that even the centers of today's massive ETGs have not evolved passively. If further observations show that these two systems are typical of $z \sim 2$ quiescent galaxies, it will require a significant mixing of stars into the centers of massive galaxies over time. This cannot be explained solely by inside-out growth, a paradigm that is supported largely by the observed modest evolution in the central density compared to the dramatic evolution seen in the outskirts of massive galaxies (e.g., \citealt{Shannon2013}). Instead these first two observations suggest that the central chemical abundances evolve substantially, which implies that although mergers may have the largest effect on the galaxy outskirts, they are nonetheless able to pollute the galaxy centers significantly. More work is needed to address the question of whether cosmologically motivated merger histories can explain the apparent chemical evolution.

The $\alpha$-enhancement is a common method of estimating formation timescale (t$_{f}$) of low-redshift quiescent galaxies, for example, using the \cite{Thomas2005} relation: [$\alpha$/Fe] $\approx$ $\frac{1}{5}-\frac{1}{6} \log(t_{f}$). Applying this linear relation to MRG-M0138 results in unrealistically short formation timescale of $\sim$ 0.05 Gyr. This suggests that the assumptions in the \cite{Thomas2005} model, such as closed box chemical evolution, a Salpeter slope for the initial mass function (IMF), or constant SFHs, are too simplistic to be able to explain these observations. Alternatively the adopted SNe yields may not be valid, particularly at high metallicity. Also, t$_{f}$ may depend not only on [$\alpha$/Fe], but also on other parameters such as total metallicity (e.g., \citealt{Yan2019}). 

Moreover, for the case of MRG-M0138, the high $\alpha$-enhancement is observed simultaneously with a high [Fe/H]. It is a long-standing problem to reproduce both high metallicities and high [$\alpha$/Fe] simultaneously in models, as simply truncating the star formation (e.g., with AGN feedback) produces a high [$\alpha$/Fe] but a low [Fe/H] (e.g., \citealt{Okamoto,DeLucia2017}). MRG-M0138 makes this problem even worse, as it has even higher [Fe/H] and [$\alpha$/Fe] than local galaxies. Simple chemical evolution models that only vary the star-formation timescale, such as the line shown in the lower panel of Figure~4, cannot match our measurements. One possible solution to this problem, though not necessarily a unique one, is to alter the initial mass function (IMF): a top-heavy IMF results in more massive stars and therefore more $\alpha$-elements.

The possible need for a top-heavy IMF in high-$z$ dusty star-forming galaxies has been suggested in some observations and models (e.g., \citealt{Zhang2018Natur,Cai2020_IMF}).  
For example, \cite{Zhang2018Natur} inferred a top-heavy IMF based on observations of isotopologue ratios of CO in sub-millimeter galaxies at \textit{z} $\sim$ 2-3. These dusty, highly star-forming galaxies are plausible progenitors of massive quiescent galaxies at \textit{z} $\sim$ 2 like MRG-M0138. However, the CO ratios are sensitive to the rotation speeds of massive stars, which depend on metallicity and possibly on ISM properties in ways that are not well understood \citep{Romano19}. The stellar abundance pattern we have measured may provide a complementary diagnostic of the high-mass IMF, but detailed galactic chemical evolution modeling is needed to derive such constraints and assess their robustness. We plan to perform such modeling in future work.

In summary, MRG-M0138 is a massive, disk-dominated quiescent galaxy with a formation redshift of \textit{z$_{\rm form}$} $\sim$ 3.3. The star formation history of this galaxy seems to have been very short across the whole disk, based on the uniformly high [Mg/Fe], though star formation may have begun earlier in the central regions based on our marginal detection of a negative [Fe/H] gradient. Quantitatively reproducing the high [Mg/Fe] of this galaxy, especially coupled with its high [Fe/H], is challenging and may require alterations to standard galactic chemical evolution models such as a top-heavy IMF. The fact that MGR-M0138 has distinct chemical abundances from local analogs and also from its only well-studied counterpart at \textit{z} $\sim$ 2, COSMOS-11494, motivates the need for deep spectroscopy of a larger sample of high-\textit{z} quiescent galaxies, which will become feasible with next-generation facilities such such as \emph{JWST} and 30-m class telescopes.

\acknowledgments

The authors thank Charlie Conroy for his assistance with the spectral fitting code, \textit{alf}, and Jenny Greene for providing relevant MASSIVE survey data. 

Part of the data presented herein were obtained at the W. M. Keck Observatory, which is operated as a scientific partnership among the California Institute of Technology, the University of California and the National Aeronautics and Space Administration. The Observatory was made possible by the generous financial support of the W. M. Keck Foundation. 

This paper includes data gathered with the 6.5 meter Magellan Telescopes located at Las Campanas Observatory, Chile.

RSE acknowledges funding from the European Research Council (ERC) under the 
European Union's Horizon 2020 research and innovation programme (grant agreement No 669253)

The authors wish to recognize and acknowledge the very significant cultural role and reverence that the summit of Maunakea has always had within the indigenous Hawaiian community.  We are most fortunate to have the opportunity to conduct observations from this mountain.

\bibliography{sample63}{}

\bibliographystyle{aasjournal}

\end{document}